Phys. Lett. B 873 (2026) 140134

Contents lists available at ScienceDirect

## Physics Letters B

journal homepage: www.elsevier.com/locate/physletb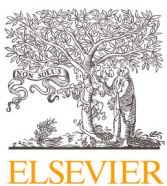
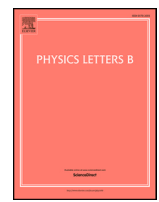

Letter

# Exploring the universe expansion history with f(R,T) gravity: Constraints on cosmological parameters

Mustapha Lamaaoune

*Team of Modern and Applied Physics, Laboratory of Research in Physics and Engineering Sciences, Sultan Moulay Slimane University, Polydisciplinary Faculty, Beni Mellal, Morocco*A R T I C L E   I N F O

Editor: P Brax

Parameter estimation
MCMC
Observational data
Deceleration parameter
Energy condition
Observational constraintsA B S T R A C T

This work examines the cosmological implications of two functional forms of $f(R,T) = R + \alpha T^n$ gravity: for two different value of $n$ where $n = 1$ and $n \neq 1$, and $\alpha$ and $n$ are free parameters. The modified Friedmann equations are derived, and the cosmic evolution of the Hubble parameter $H(z)$ is determined. Cosmological parameters are estimated through $\chi^2$ minimization and MCMC analysis using the emcee algorithm, with model parameters constrained by various observational datasets. Both models reproduce late-time acceleration and remain observationally indistinguishable from $\Lambda$CDM, while allowing small deviations parameterized by $\alpha$ and $n$. The cosmological behavior of the deceleration parameter $q(z)$, the jerk, the snap parameter $s(z)$, and the effective equation of state parameter $\omega$ is also analyzed. The results indicate that the Universe transitioned from deceleration to late-time accelerated expansion, consistent with the $\Lambda$CDM model. Analysis of the energy conditions reveals that the NEC and DEC are satisfied, while the SEC is violated, which explains the transition from a decelerating matter-dominated epoch to the present accelerated phase. These findings indicate that the proposed $f(R,T)$ models are compatible with current observational data and provide a viable alternative to $\Lambda$CDM in describing cosmic acceleration.## 1. Introduction

The discovery of the accelerated expansion of the Universe at the end of the 20th century has revealed a profound challenge in modern cosmology [1–7]. Subsequent observations of type Ia supernovae (SNe Ia) confirm that the Universe is expanding at an accelerated rate [8,9]. Within the framework of general relativity (GR), this late-time acceleration is typically attributed to a mysterious dark energy component, most often modeled as a cosmological constant $\Lambda$. Although the standard $\Lambda$ Cold Dark Matter ($\Lambda$CDM) model successfully describes a wide range of cosmological observations, it faces persistent theoretical challenges, including the fine-tuning and coincidence problems [10,11]. The physical nature of dark energy remains elusive, motivating extensive exploration of extensions or alternatives to general relativity that could explain cosmic acceleration without invoking exotic matter components. This context encourages the investigation of alternative models for the accelerated expansion of the Universe [12–14].

Among the wide class of modified gravity theories, $f(R)$ gravity extends the Einstein-Hilbert action to a general function of the Ricci scalar $R$ [15–21]. Related extensions, including $f(G)$ and $f(R,G)$ gravity, where $G$ is the Gauss-Bonnet invariant, have also been widely investigated. These models provide a unified theoretical framework that can describe both the early inflationary epoch and the late-time acceleration of the Universe. A further generalization, $f(R,T)$ gravity, introduces a gravitational Lagrangian that depends on both the Ricci scalar and the trace of the energy-momentum tensor [22–28]. The explicit dependence on $T$ can arise from quantum effects, exotic imperfect fluids, or couplings between matter and geometry. This leads to dynamics more complex than those in pure $f(R)$ gravity, in $f(R,T)$ theories, the covariant divergence of the stress-energy tensor is generally non-zero. This results in non-geodesic motion and an additional force acting on matter. This property positions $f(R,T)$ models as a promising framework for exploring potential deviations from general relativity at cosmological scales [29–31]. Compared with pure $f(R)$ models, the dependence on $T$ introduces direct matter-geometry coupling. This coupling can be the source of effective late-time acceleration without an explicit $\Lambda$ and modifies the effective equation of state and the growth of perturbations. These features motivate confronting $f(R,T)$ predictions with data. However, the additional degrees of freedom also require stronger observational constraints to avoid degeneracy with $\Lambda$CDM.

In this work we consider the modified gravity $f(R,T) = R + \alpha T^n$ by taking the following case the general nonlinear sector $n \neq 1$ and the special case $n = 1$, which is not a distinct theory but simply a particular limit of the nonlinear class. These models introduce two additional degrees

*E-mail address:* lamaaoune2944massi@gmail.com

https://doi.org/10.1016/j.physletb.2025.140134
Received 17 September 2025; Received in revised form 5 December 2025; Accepted 22 December 2025
Available online 27 December 20250370-2693/© 2025 The Author(s). Published by Elsevier B.V. Funded by SCOAP³. This is an open access article under the CC BY license (http://creativecommons.org/licenses/by/4.0/).

of freedom beyond standard ΛCDM, the coupling constant $\alpha$ and the power-law index $n$. These parameters together characterize the strength and form of the matter-geometry interaction. For clarity and consistency, the analysis is performed in a flat Universe dominated by matter and dark energy. The density parameters are $\Omega_k = 0$, $\Omega_m$, and $\Omega_\Lambda$, allowing a direct comparison with ΛCDM [10,11]. Explaining the observed cosmic acceleration remains one of the central challenges in cosmology. While ΛCDM provides an excellent fit to current observations, it relies on a cosmological constant whose theoretical origin and fine-tuning problems remain unresolved. Modified gravity theories, such as $f(R,T)$ gravity, offer an alternative by allowing gravitational dynamics to depend directly on both spacetime curvature and the matter content. Exploring the linear case $n = 1$ alongside its nonlinear extension $n \neq 1$ enables us to determine how different forms of matter-geometry coupling influence the expansion history and the cosmological parameters $\Omega_m$ and $\Omega_\Lambda$. By confronting these models with high-precision datasets-including cosmic chronometers, Type Ia supernovae, and baryon acoustic oscillations-and by examining cosmographic indicators and energy conditions, we assess their viability and gain insight into the physical mechanisms that may drive the accelerated expansion of the Universe [12–14].

Constraining the parameters of these models using [32] with the help of the emcee library [33] and precise cosmological data is essential for evaluating their viability. Advances in observational cosmology now allow for increasingly accurate tests of modified gravity models. Cosmic chronometers provide model-independent measurements of the Hubble parameter $H(z)$, while the Pantheon+SH0ES compilation of Type Ia supernovae offers the most current dataset for distance modulus measurements [34–41]. When combined with baryon acoustic oscillation (BAO) data, these datasets impose stringent constraints on the Hubble constant $H_0$, the matter density $\Omega_m$, the dark energy density $\Omega_\Lambda$, and the modified gravity parameters $\alpha$ and $n$.

This work is organized as follows: Section 2 provides a brief review of the theoretical framework of $f(R,T)$ gravity; Section 3 details the observational datasets and the statistical methodology used for parameter estimation of the specific model forms. Section 4 presents results on parameter constraints, cosmological diagnostics, and the behavior of the energy conditions. Section 5 summarizes the findings and discusses their implications for the viability of $f(R,T)$ gravity as an alternative to the standard ΛCDM cosmology.

## 2. $f(R,T)$ gravity model

We consider the modified gravity theory based on the $f(R,T)$ action, given in [22] by

$$S = \frac{1}{2\kappa^2} \int d^4x \sqrt{-g} f(R,T) + \int d^4x \sqrt{-g} \mathcal{L}_m, \quad (2.1)$$

Here, $\kappa^2 = 8\pi G$, $R$ denotes the Ricci scalar, $g$ is the determinant of the metric $g_{\mu\nu}$, and $T = g^{\mu\nu} T_{\mu\nu}$ represents the trace of the energy-momentum tensor. The function $f(R,T)$ is an arbitrary function of both $R$ and $T$, while $\mathcal{L}_m$ denotes the matter Lagrangian density.

The energy-momentum tensor is defined by

$$T_{\mu\nu} = -\frac{2}{\sqrt{-g}} \frac{\delta(\sqrt{-g}\mathcal{L}_m)}{\delta g^{\mu\nu}}. \quad (2.2)$$

Varying the action (2.1) with respect to the metric $g^{\mu\nu}$ yields the modified field equations [22–26]

$$f_R R_{\mu\nu} - \frac{1}{2} g_{\mu\nu} f(R,T) + (g_{\mu\nu} \Box - \nabla_\mu \nabla_\nu) f_R = \kappa^2 T_{\mu\nu} - f_T (T_{\mu\nu} + \Theta_{\mu\nu}), \quad (2.3)$$

In these equations

$$f_R \equiv \frac{\partial f}{\partial R}, \quad f_T \equiv \frac{\partial f}{\partial T}, \quad (2.4)$$

and the variation of the trace $T$ with respect to the metric is given by

$$\delta T = T_{\mu\nu} \delta g^{\mu\nu} + g^{\mu\nu} \delta T_{\mu\nu}, \quad (2.5)$$

with

$$\Theta_{\mu\nu} \equiv g^{\alpha\beta} \frac{\delta T_{\alpha\beta}}{\delta g^{\mu\nu}}. \quad (2.6)$$

This leads to the following expression

$$\delta T = (T_{\mu\nu} + \Theta_{\mu\nu}) \delta g^{\mu\nu}. \quad (2.7)$$

In this analysis, we assume the energy-momentum tensor describes a perfect fluid [22,82–84].

$$T_{\mu\nu} = (\rho + p) u_\mu u_\nu + p g_{\mu\nu}, \quad (2.8)$$

Here, $\rho$ denotes the energy density, $p$ the pressure, and $u^\mu$ the four-velocity of the fluid.

The cosmological implications are examined under the assumption of a spatially flat, isotropic, and homogeneous Universe, described by the Friedmann-Lemaître-Robertson-Walker (FLRW) metric [42–46]. The FLRW metric is given by

$$ds^2 = -dt^2 + a(t)^2 (dx^2 + dy^2 + dz^2), \quad (2.9)$$

where $a(t)$ denotes the cosmic scale factor and $t$ represents the cosmic time.

Eq. (2.3) can be recast in the form of a modified Einstein equation, $G_{\mu\nu} = \kappa^2 T_{\mu\nu}^{\text{eff}}$, where $T_{\mu\nu}^{\text{eff}}$ denotes an effective energy-momentum tensor [22–28]. From (2.3), it follows that

$$T_{\mu\nu}^{\text{eff}} = \frac{1}{\kappa^2} \left[ \kappa^2 T_{\mu\nu} - f_T (T_{\mu\nu} + \Theta_{\mu\nu}) + \frac{1}{2} f(T) g_{\mu\nu} \right]. \quad (2.10)$$

The trace of this effective energy-momentum tensor is given by

$$T = g^{\mu\nu} T_{\mu\nu} = -\rho + 3p. \quad (2.11)$$

For a perfect fluid, the following relation holds

$$\Theta_{\mu\nu} = -2 T_{\mu\nu} + p g_{\mu\nu}, \quad (2.12)$$

This result leads to

$$T_{\mu\nu} + \Theta_{\mu\nu} = -T_{\mu\nu} + p g_{\mu\nu}. \quad (2.13)$$

The late-time Universe, dominated by matter and dark energy, is now considered. The total energy density and pressure are given by

$$\rho = \rho_m + \rho_\Lambda, \quad p = p_m + p_\Lambda = 0 - \rho_\Lambda = -\rho_\Lambda, \quad (2.14)$$

Here, $\rho_m$ denotes the energy density of pressureless matter ($p_m = 0$), and $\rho_\Lambda$ is the constant energy density of dark energy ($p_\Lambda = -\rho_\Lambda$).

The trace of the energy-momentum tensor $T$ then becomes

$$T = -\rho + 3p = -(\rho_m + \rho_\Lambda) + 3(-\rho_\Lambda) = -\rho_m - 4\rho_\Lambda. \quad (2.15)$$

The evolution of the matter density with redshift $z$ is given by

$$\rho_m = \rho_m (1+z)^3, \quad (2.16)$$

where $\rho_m$ denotes the present matter density of the Universe and $\rho_\Lambda$ remains constant.

The present-day critical density $\rho_c$ and density parameters $\Omega_m$ and $\Omega_\Lambda$ are introduced as follows:

$$\rho_c = \frac{3H_0^2}{\kappa^2}, \quad \Omega_m = \frac{\rho_m}{\rho_c}, \quad \Omega_\Lambda = \frac{\rho_\Lambda}{\rho_c}, \quad (2.17)$$

where $H_0$ is the present-day Hubble constant, which typically varies in the range 68 to 73 km s$^{-1}$ $Mpc^{-1}$.

In terms of these parameters and redshift $z$, the densities and traces are

$$\rho = \rho_c \left[ \Omega_m (1+z)^3 + \Omega_\Lambda \right],$$
$$p = -\rho_c \Omega_\Lambda,$$
$$T = -\rho_c \Omega_m (1+z)^3 - 4\rho_c \Omega_\Lambda. \quad (2.18)$$

In this study, we consider the modified gravity function $f(R,T) = R + \alpha T^n$, where $\alpha$ and $n$ are free parameter, this model is chosen because it provides a simple and controlled extension of General Relativity while keeping the standard $R$ term intact. The power-law term in $T^n$ term introduces nonlinear matter-geometry coupling, allowing one to test subtle deviations from $\Lambda$CDM without over-parameterizing the theory. The model remains theoretically tractable and easily constrained by current cosmological observations for two extra parameters. Taking into account the expression of the modified gravity $f(R,T)$ function, the Eq. (2.3) became

$$R_{\mu\nu} - \frac{1}{2} R g_{\mu\nu} = \kappa^2 T_{\mu\nu} + \alpha n T^{n-1}(-T_{\mu\nu} + p g_{\mu\nu}) + \frac{\alpha}{2} T^n g_{\mu\nu}. \quad (2.19)$$

Substituting Eqs. (2.15) and (2.18) into (2.19), the Hubble parameter is obtained as

$$H(z)^2 = H_0^2 \Bigg[ \Omega_m(1+z)^3 + \Omega_\Lambda$$
$$+ \frac{\alpha}{6H_0^2} \left( \frac{3H_0^2}{\kappa^2} [\Omega_m(1+z)^3 + 4\Omega_\Lambda] \right)^n$$
$$+ \frac{\alpha n}{3H_0^2 \kappa^2} \left( \frac{3H_0^2}{\kappa^2} [\Omega_m(1+z)^3 + 4\Omega_\Lambda] \right)^{n-1} \frac{3H_0^2}{\kappa^2} \Omega_m(1+z)^3 \Bigg]. \quad (2.20)$$

The effective energy density and pressure are defined by the following expressions

$$\rho = \frac{3H_0^2}{\kappa^2}[\Omega_m(1+z)^3 + \Omega_\Lambda] + \frac{\alpha n}{\kappa^2} \left( \frac{3H_0^2}{\kappa^2} \Omega_m(1+z)^3 \right) \left( -\frac{3H_0^2}{\kappa^2}[\Omega_m(1+z)^3 + 4\Omega_\Lambda] \right)^{n-1}$$
$$- \frac{\alpha}{2\kappa^2} \left( -\frac{3H_0^2}{\kappa^2}[\Omega_m(1+z)^3 + 4\Omega_\Lambda] \right)^n,$$
$$p = -\frac{3H_0^2}{\kappa^2}\Omega_\Lambda + \frac{\alpha}{2\kappa^2}\left(-\frac{3H_0^2}{\kappa^2}[\Omega_m(1+z)^3 + 4\Omega_\Lambda]\right)^n. \quad (2.21)$$

In the following sections, our goal is to validate our model for both cases: the linear case $n = 1$ and the nonlinear case $n \neq 1$. We use various cosmological datasets, including CC, Pantheon$^+$SH0ES, their joint analysis of CC+Pantheon$^+$SH0ES, and CC+Pantheon$^+$SH0ES+BAO, to constrain the different parameter $H_0$, $\Omega_m$, $\Omega_\Lambda$, $\alpha$ and $n$ and validate the different cosmological parameter and energy condition.

## 3. Observational constraints

In this section, we present the observational datasets used in our analysis, namely Cosmic Chronometers, Pantheon$^+$SH0ES, and BAO. These datasets are employed to constrain the key cosmological parameters of the model, including the Hubble constant $H_0$, the dark energy density $\Omega_\Lambda$, the matter density $\Omega_m$, and the model parameters $\alpha$ and $n$ that characterize the modified Hubble function $H(z)$. To explore the parameter space efficiently and obtain reliable estimates of their posterior distributions, we perform a Markov Chain Monte Carlo (MCMC) analysis [32]. The sampling is carried out using the *emcee* python package [33], which implements an affine-invariant ensemble sampler well suited for multidimensional parameter estimation. The likelihood function is constructed under the assumption of Gaussian observational uncertainties, leading to a probability distribution of the form

$$\mathcal{L} \propto \exp\left(-\frac{\chi^2}{2}\right), \quad (3.1)$$

$\chi^2$ represents the total chi-square statistic, quantifying the difference between the theoretical model and the observed data. The parameter space is analyzed using flat, uniform priors defined over the ranges $60 < H_0 < 80$, $-0.5 < \Omega_\Lambda < 1.5$, $0 < \Omega_m < 1$, $-3 < \alpha < 3$, and $-3 < n < 3$. Combinations of datasets are considered, including cosmic chronometers (CC, 31 points), the Pantheon$^+$SH0ES supernova compilation (1701 data points), and baryon acoustic oscillation (BAO) measurements from various surveys. Joint analyses are also conducted using combinations such as CC+Pantheon$^+$SH0ES and CC+Pantheon$^+$SH0ES+BAO [34–41,47]. For selected dataset combinations, two-dimensional posterior contour plots are presented, showing $1\sigma$ and $2\sigma$ confidence levels to illustrate the credibility and parameter correlations in the proposed model.

**Table 1**
Cosmic chronometer measurements of the Hubble parameter $H(z)$ with uncertainties and corresponding references.

| $z$ | $H_{obs}$ (km/s/Mpc) | $\sigma_H$ (km/s/Mpc) | Reference |
|---|---|---|---|
| 0.07 | 69.0 | 19.6 | [48] |
| 0.09 | 69.0 | 12.0 | [49] |
| 0.12 | 68.6 | 26.2 | [50] |
| 0.17 | 83.0 | 8.0 | [51] |
| 0.179 | 75.0 | 4.0 | [51] |
| 0.199 | 75.0 | 5.0 | [51] |
| 0.20 | 72.9 | 29.6 | [49] |
| 0.27 | 77.0 | 14.0 | [50] |
| 0.28 | 88.8 | 36.6 | [51] |
| 0.352 | 83.0 | 14.0 | [51] |
| 0.3802 | 83.0 | 13.5 | [53] |
| 0.40 | 95.0 | 17.0 | [50] |
| 0.4004 | 77.0 | 10.2 | [53] |
| 0.4247 | 87.1 | 11.2 | [53] |
| 0.4497 | 92.8 | 12.9 | [53] |
| 0.47 | 89.0 | 23.0 | [54] |
| 0.4783 | 80.9 | 9.0 | [53] |
| 0.48 | 97.0 | 62.0 | [50] |
| 0.593 | 104.0 | 13.0 | [52] |
| 0.68 | 92.0 | 8.0 | [52] |
| 0.781 | 105.0 | 12.0 | [52] |
| 0.875 | 125.0 | 17.0 | [52] |
| 0.88 | 90.0 | 40.0 | [50] |
| 0.90 | 117.0 | 23.0 | [50] |
| 1.037 | 154.0 | 20.0 | [52] |
| 1.30 | 168.0 | 17.0 | [52] |
| 1.363 | 160.0 | 33.4 | [48] |
| 1.43 | 177.0 | 18.0 | [52] |
| 1.53 | 140.0 | 14.0 | [52] |
| 1.75 | 202.0 | 40.0 | [52] |
| 1.965 | 186.5 | 50.4 | [52] |

### 3.1. Cosmic chronometer

The cosmic chronometer (CC) approach provides a robust and model-independent method for constraining the expansion history of the Universe. Unlike techniques that rely on specific cosmological assumptions, the CC method employs differential dating of passively evolving galaxies to directly determine the Hubble parameter $H(z)$ at various redshifts. This analysis utilizes a compilation of 31 observational $H(z)$ data points spanning the redshift range $0.07 \leq z \leq 1.965$ [47]. These measurements serve as an independent probe of the cosmic expansion rate and enable a rigorous evaluation of the theoretical model using observational data [48–54].

The Hubble parameter is defined as $H(z) = -\frac{1}{1+z}\frac{dz}{dt}$, which quantifies the rate of change of redshift with respect to cosmic time. Model predictions are compared with observational data using the chi-square function. This statistical approach facilitates the estimation of best-fit values for the model parameters, as demonstrated below.

$$\chi^2 = \sum_{i=1}^{31} \left[ \frac{H_{th}(z_i|\Theta) - H_{obs}(z_i)}{\sigma_H(z_i)} \right]^2 \quad (3.2)$$

Here, $H(\Theta, z_i)$ denotes the theoretical Hubble parameter at redshift $z_i$, and $H_{obs}(z_i)$ represents the observed value. The term $\sigma_H$ indicates the standard error associated with the measured values of $H_{obs}(z_i)$ in the cosmological model Table 1 [75–81,85,86].

### 3.2. Pantheon$^+$SH0ES dataset

Type Ia supernovae (SNeIa) are reliable cosmic distance indicators because of their consistent intrinsic brightness, which makes them essential for investigating the expansion of the Universe. The Pantheon$^+$ dataset comprises 1071 apparent magnitude measurements for 1550 SNeIa across a redshift range of $z = 0.001$ to 2.26 [34,35].
In this dataset, the observed distance modulus $\mu$ is directly related to the luminosity distance $D_L(z)$ as follows [55,56]

$$\mu_{th}(z) = m_b - M_B = 5\log\left(\frac{D_L(z)}{1\text{Mpc}}\right) + 25, \tag{3.3}$$

$m_b$ denotes the apparent magnitude, and $M_B$ is the absolute magnitude of a standard source.

The luminosity distance is defined by the following expression

$$D_L(z) = \frac{c(1+z)}{H_0} S_k\left(H_0 \int_0^z \frac{dz^*}{H(z^*)}\right), \tag{3.4}$$

In this equation,

$$S_k(x) = \begin{cases} \sinh(x\sqrt{\Omega_k})/\Omega_k, & \Omega_k > 0 \\ x, & \Omega_k = 0 \\ \sin(x\sqrt{|\Omega_k|})/|\Omega_k|, & \Omega_k < 0 \end{cases} \tag{3.5}$$

$c$ denotes the speed of light. For a flat Universe with $\Omega_k = 0$, the luminosity distance simplifies as follows [34,35,55]

$$D_L(z) = c(1+z)\int_0^z \frac{dz'}{H(z')}. \tag{3.6}$$

The model is evaluated by comparing predicted and observed distance moduli. The resulting residuals are subsequently used to calculate the chi-squared statistic [34,35,55,56].

$$\chi^2_{pan^+S} = \Delta\mu^T C^{-1}_{pan^+S} \Delta\mu, \tag{3.7}$$

Here, $\Delta\mu = \mu_{th} - \mu_{obs}$, where $\mu_{th}$ is the model-predicted distance modulus and $\mu_{obs}$ is the observed value from the Pantheon$^+$SH0ES dataset. $C^{-1}_{SNe}$ denotes the covariance matrix that incorporates both statistical and systematic uncertainties.

### 3.3. BAO datasets

Baryon Acoustic Oscillations (BAO) began in the early Universe when ordinary matter and photons were closely linked by Thomson scattering. This connection caused oscillations in the matter-photon fluid, as strong radiation pressure prevented gravitational collapse from occurring. Today, we observe these oscillations as BAO. The typical scale is set by the sound horizon, $r_s = 146$ Mpc, measured at the photon decoupling epoch $z^* = 1090$ for this study [57].
This analysis uses BAO distance data from several surveys, like SDSS(R) [58], the 6dF Galaxy Survey [59], BOSS CMASS [60], and three separate results from the WiggleZ survey [61–64]. The BAO data help us constrain cosmological models using a distance-redshift ratio:

$$d_z = \frac{r_s(z^*)}{D_v(z)}, \tag{3.8}$$

where $r_s(z^*)$ is the sound horizon at the time when light separated from matter [65], calculated as:

$$r_s(a) = \int_0^a \frac{c_s da}{a^2 H(a)}. \tag{3.9}$$

We define the dilation scale $D_v(z)$ as

$$D_v(z) = \left(z d_A^2(z) D_H(z)\right)^{1/3}, \tag{3.10}$$

with $d_A(z)$ being the angular diameter distance. For a flat Universe

$$d_A(z^*) = c\int_0^{z^*} \frac{dz'}{H(z')}. \tag{3.11}$$

The $\chi^2$ statistic for BAO compares the predicted and observed values [66]

$$\chi^2_{BAO} = X^T C^{-1} X, \tag{3.12}$$

where $X$ is a column measuring the differences

$$X = \begin{bmatrix} \frac{d_A(z^*)}{D_v(0.106)} - 30.84 \\ \frac{d_A(z^*)}{D_v(0.35)} - 10.33 \\ \frac{d_A(z^*)}{D_v(0.57)} - 6.72 \\ \frac{d_A(z^*)}{D_v(0.44)} - 8.41 \\ \frac{d_A(z^*)}{D_v(0.6)} - 6.66 \\ \frac{d_A(z^*)}{D_v(0.73)} - 5.43 \end{bmatrix}, \tag{3.13}$$

and $C^{-1}$ is the inverse covariance matrix [66]

$$C^{-1} = \begin{bmatrix} 0.52552 & -0.03548 & -0.07733 & -0.00167 & -0.00532 & -0.0059 \\ -0.03548 & 24.9707 & -1.25461 & -0.02704 & -0.08633 & -0.09579 \\ -0.07733 & -1.25461 & 82.9295 & -0.05895 & -0.18819 & -0.20881 \\ -0.00167 & -0.02704 & -0.05895 & 2.9115 & -2.98873 & 1.43206 \\ -0.00532 & -0.08633 & -0.18819 & -2.98873 & 15.9683 & -7.70636 \\ -0.0059 & -0.09579 & -0.20881 & 1.43206 & -7.70636 & 15.2814 \end{bmatrix}. \tag{3.14}$$

### 3.4. Joint dataset

We estimate the parameters by combining the Hubble, Pantheon$^+$SH0ES, and BAO datasets. Their joint likelihood function is used. In the first case, we combine the Cosmic Chronometer and Pantheon$^+$SH0ES datasets [34,35,47]. The likelihood function $\chi^2_{OHD+P^+S}$ is defined as follows

$$\chi^2_{CC+P^+S} = \chi^2_{CC} + \chi^2_{P^+S} \tag{3.15}$$

Alternatively, the three datasets-Cosmic Chronometer, Pantheon$^+$SH0ES, and BAO [34–41,47] are combined to construct the total joint likelihood function.

$$\chi^2_{CC+P^+S+BAO} = \chi^2_{CC} + \chi^2_{P^+S} + \chi_{BAO} \tag{3.16}$$

## 4. Constraints on the models from fitting analysis

In this section, we present the results of the model fitting analysis. We minimize the corresponding $\chi^2$ function using Markov Chain Monte Carlo (MCMC) sampling with the emcee library [32,33], allowing for a thorough exploration of the parameter space and the identification of statistical distributions consistent with the observational data. We report best-fit values for the model parameters using CC, Pantheon$^+$SH0ES, and their joint analyses with BAO [34–41,47]. The results are summarized in Tables 3 and 2.

### 4.1. Constraints on $n \neq 1$

The model is fitted to observational data using the $\chi^2$ minimization method to determine the best-fit values of the free parameters. Numerical results are presented in Table 2. Fig. 1 presents the corner plot, which displays the one-dimensional marginalized posterior distributions and two-dimensional confidence contours for the datasets (CC, Pantheon$^+$SH0ES, CC + Pantheon$^+$SH0ES, and CC + Pantheon$^+$SH0ES + BAO), generated with the GetDist library [67]. The confidence contours for $H_0$, $\Omega_m$, $\Omega_\Lambda$, $\alpha$ and $n$ are summarized in Table 2.
The parameter constraints summarized in Table 2 confirm that the $f(R,T) = R + \alpha T^n$ model with $n \neq 1$ remains statistically consistent with the standard $\Lambda$CDM cosmology across all dataset combinations [10,11]. The inferred Hubble constant is remarkably stable among all fits, lying in the narrow interval $H_0 \simeq 67.2 - 67.9$ kms$^{-1}$Mpc$^{-1}$, which closely matches the Planck 2018 determination and disfavors the higher $H_0$ values implied by local SH0ES measurements [34,35]. This suggests that, within this modified gravity framework, late-time cosmic expansion behaves in a manner largely compatible with early-Universe constraints.

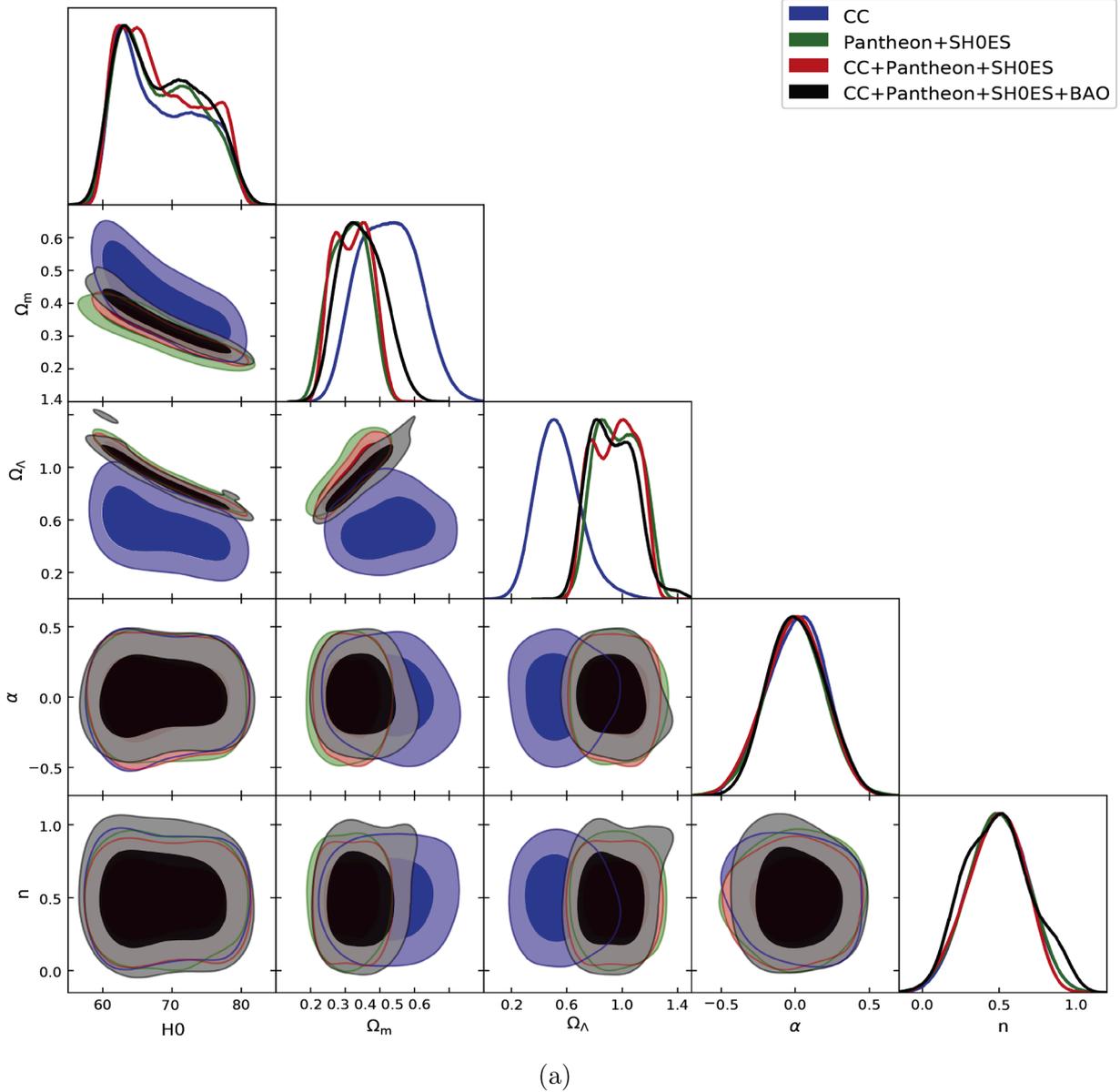

(a)

**Fig. 1.** The likelihood contours at $1\sigma$ and $2\sigma$ confidence levels (CLs) and the 1D posterior distributions for CC, Pantheon+SH0ES, the joint analysis of CC + Pantheon+SH0ES and CC + Pantheon+SH0ES + BAO.

The matter density parameter $\Omega_m$ exhibits a clear dependence on the dataset selection. CC data alone yield a relatively high value, $\Omega_m \approx 0.43$, whereas Pantheon+SH0ES and the combined datasets constrain it to values much closer to the canonical $\Omega_m \approx 0.3$ including BAO data further stabilizes this parameter while slightly reducing its uncertainty [34–41]. Moreover, the corresponding dark energy density $\Omega_\Lambda$ reflects this shift, remaining compatible with $\Omega_\Lambda \approx 0.7$ once SN and BAO data are included. The unusually large value obtained from Pantheon+SH0ES alone arises from its strong preference for lower $\Omega_m$, but remains consistent within statistical uncertainties.

Crucially, the modified gravity parameters show no significant evidence for deviations from general relativity. The coupling parameter $\alpha$ is fully compatible with zero at the $2\sigma$ level for every dataset combination, indicating no statistically required matter-geometry coupling in current observations. Similarly, the exponent $n$ remains tightly constrained around $n \approx 0.5$, with uncertainties $\mathcal{O}(0.2)$, but does not deviate sufficiently from unity to demand strong nonlinear modifications of the matter trace contribution.

Overall, these results demonstrate that although the $f(R, T)$ gravity framework with $n \neq 1$ can successfully reproduce the observed expansion history, current data do not favor departures from general relativity or the $\Lambda$CDM paradigm [10,11].

*4.2. Constraints on $n = 1$*

We fit the model to the observational data using $\chi^2$ minimization, obtaining the best-fit values for its free parameters based on the considered data sets. Table 3 presents the numerical results, illustrating the comparison of the fitted parameters across different datasets. Fig. 2 displays the corner plot with one-dimensional marginalized posterior distributions and two-dimensional confidence contours for each parameter, generated with the GetDist library [67]. The specific constraints on $H_0$, $\Omega_m$, $\Omega_\Lambda$, and $\alpha$, as inferred from the data, are detailed in Table 3.

The Analysis of case where $n = 1$ model using multiple observational datasets yields results consistent with the concordance $\Lambda$CDM cosmology [10,11]. To begin, for the cosmic chronometer (CC) data

**Table 2**

The table represents the best-fit values of the parameters for $n \neq 1$ and for various datasets CC, Pantheon$^+$SH0ES, and joint datasets CC + Pantheon$^+$SH0ES and CC + Pantheon$^+$SH0ES + BAO.

|  | CC | Pantheon$^+$SH0ES | CC + Pantheon$^+$SH0ES | CC + Pantheon$^+$SH0ES + BAO |
|---|---|---|---|---|
| $H_0$ | $67.214^{+7.898}_{-5.276}$ | $67.369^{+7.135}_{-5.375}$ | $67.387^{+8.067}_{-5.220}$ | $67.857^{+7.410}_{-5.835}$ |
| $\Omega_m$ | $0.427^{+0.084}_{-0.096}$ | $0.315^{+0.055}_{-0.065}$ | $0.324^{+0.054}_{-0.067}$ | $0.340^{+0.076}_{-0.061}$ |
| $\Omega_\Lambda$ | $0.527^{+0.158}_{-0.146}$ | $0.965^{+0.187}_{-0.177}$ | $0.958^{+0.175}_{-0.196}$ | $0.913^{+0.188}_{-0.160}$ |
| $\alpha$ | $0.024^{+0.180}_{-0.221}$ | $-0.005^{+0.187}_{-0.186}$ | $-0.002^{+0.203}_{-0.200}$ | $0.005^{+0.203}_{-0.189}$ |
| $n$ | $0.497^{+0.193}_{-0.193}$ | $0.495^{+0.200}_{-0.194}$ | $0.400^{+0.186}_{-0.201}$ | $0.485^{+0.228}_{-0.230}$ |
| $\chi^2_{\min}$ | 0.0 | −4.4 | −4.0 | −2.3 |

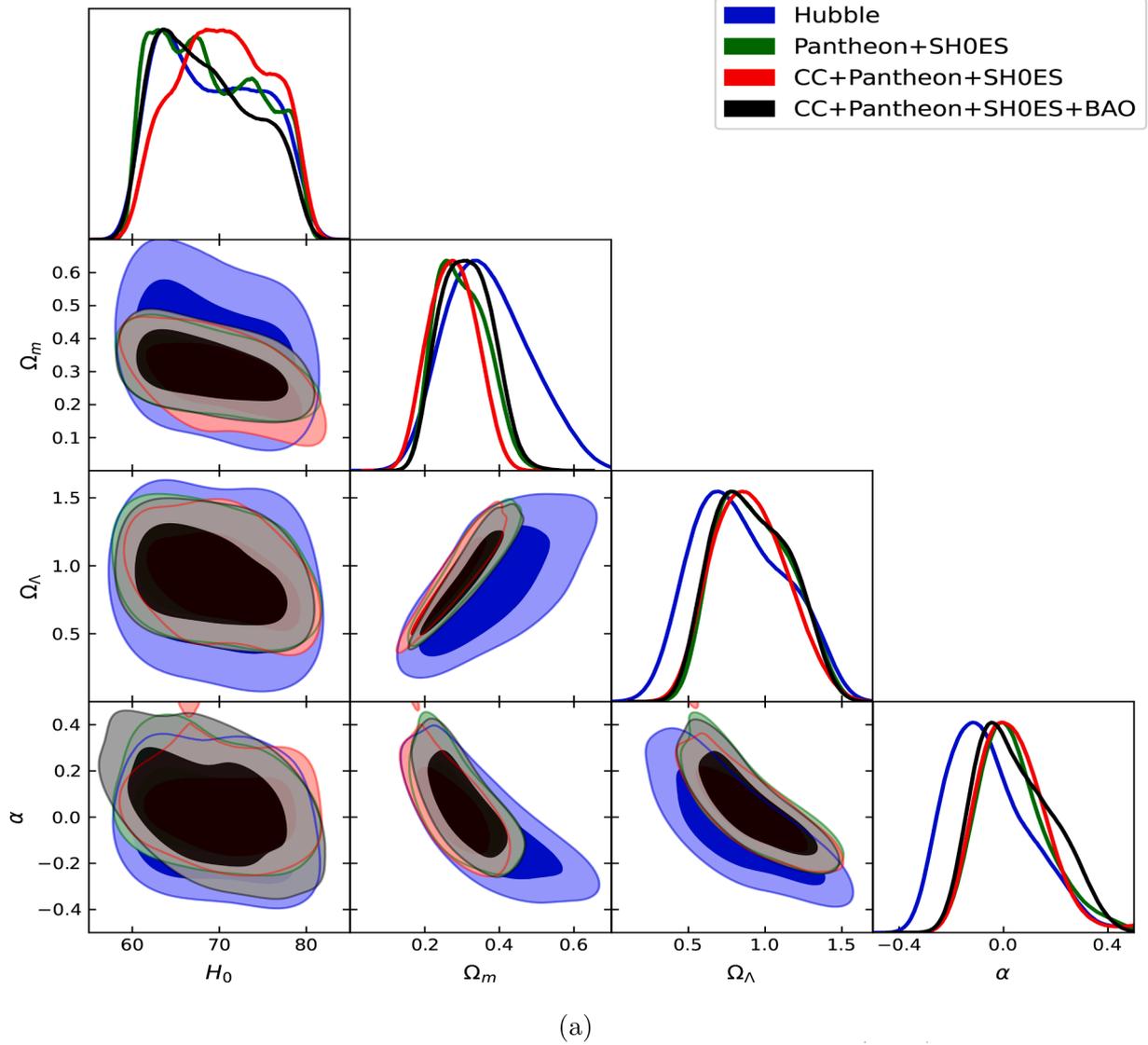

(a)

**Fig. 2.** *The likelihood contours are $1\sigma$ and $2\sigma$ confidence levels (CLs) and the 1D posterior distributions for CC, Pantheon$^+$SH0ES, the joint analysis of CC + Pantheon$^+$SH0ES, and CC + Pantheon$^+$SH0ES + BAO.*

alone, the constraint on the Hubble constant is relatively broad, $H_0 \approx 62.7 - 76.1 km/s/Mpc$, overlapping with both the early-Universe Planck 2018 determination and the late-Universe local SH0ES measurement.

The matter density parameter in this scenario is higher than the canonical $\Omega_m \simeq 0.3$, but remains compatible within the quoted errors. Moving to Pantheon$^+$SH0ES data, either alone or combined with CC, the constraints on $H_0$ become tighter and shift toward values closer to the local SH0ES determination. Specifically, the joint CC + Pantheon$^+$SH0ES analysis yields $H_0 \approx 64.6$-$76.7 km/s/Mpc$. The corresponding matter and dark energy densities remain consistent with $\Omega_m \sim 0.3$ and $\Omega_\Lambda \sim 0.7$, as expected in the standard $\Lambda$CDM scenario [10,11]. Furthermore, the inclusion of BAO data with CC and Pantheon$^+$SH0ES shifts the best-fit value of $H_0$ toward the Planck determination, yielding $H_0 \approx 62.5$-$75.0 km/s/Mpc$, and reduces parameter uncertainties compared to CC-only fits. In this combined case, the matter density parameter is tightly constrained around $\Omega_m \approx 0.3$, and the dark energy density remains compatible with the concordance model. The coupling parameter $\alpha$ remains consistent with zero within the 1!−!2$\sigma$ level for all datasets, indicating no statistically significant evidence for deviations from the standard general relativity limit within this model.

**Table 3**

The best-fit values of the parameters for the second $f(R,T)$ gravity model for various datasets CC, Pantheon$^+$SH0ES, and joint datasets CC+Pantheon$^+$SH0ES and CC+Pantheon$^+$SH0ES+BAO.

|  | CC | Pantheon$^+$SH0ES | CC+Pantheon$^+$SH0ES | CC+Pantheon$^+$SH0ES+BAO |
|---|---|---|---|---|
| $H_0$ | $68.607^{+7.495}_{-5.924}$ | $68.303^{+7.286}_{-5.626}$ | $70.450^{+6.232}_{-5.898}$ | $67.862^{+7.164}_{-5.361}$ |
| $\Omega_m$ | $0.362^{+0.124}_{-0.110}$ | $0.292^{+0.081}_{-0.063}$ | $0.275^{+0.067}_{-0.067}$ | $0.309^{+0.075}_{-0.076}$ |
| $\Omega_\Lambda$ | $0.807^{+0.381}_{-0.287}$ | $0.901^{+0.235}_{-0.298}$ | $0.887^{+0.263}_{-0.242}$ | $0.895^{+0.089}_{-0.293}$ |
| $\alpha$ | $-0.078^{+0.190}_{-0.142}$ | $0.018^{+0.156}_{-0.113}$ | $0.042^{+0.139}_{-0.119}$ | $0.019^{+0.199}_{-0.123}$ |
| $\chi^2_{\min}$ | 22.97 | 15.03 | 3425.2 | 3428.9 |

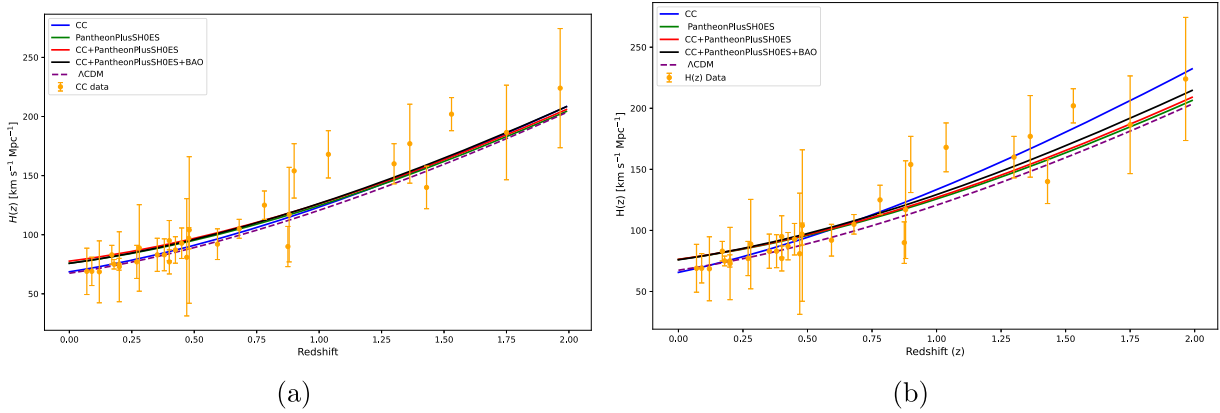

**Fig. 3.** Evolution of the Hubble parameter $H(z)$ for different value of $n$: (a) $n = 1$ and (b) $n \neq 1$. with the constraints with various observational datasets (CC, Pantheon$^+$SH0ES, CC + Pantheon$^+$SH0ES, and CC + Pantheon$^+$SH0ES + BAO). The orange points with error bars correspond to the 31-point Hubble dataset, while the purple dashed curve denotes the ΛCDM prediction with $\Omega_\Lambda = 0.7$ and $\Omega_m = 0.3$.

Overall, these results indicate that the second $f(R,T)$ model remains observationally viable and does not introduce significant tension with current cosmological data.

The modified gravity parameters remain close to general relativity limits. In particular, the coupling parameter $\alpha$ is statistically consistent with zero within $2\sigma$ level, while the exponent $n$ is constrained near 0.5 across all datasets. These results show that current observations do not require deviations from the standard ΛCDM framework. Including CC data broadens parameter ranges and increases $\Omega_m$. The evolution of the Hubble parameter $H(z)$ and distance modulus $\mu(z)$, based on best-fit parameters, is shown in Figs. 3(a) and 4 for the first and second models, respectively.

Fig. 3 presents the evolution of the Hubble parameter $H(z)$ for different $f(R,T)$ gravity models, as defined in Fig. 3(a) and (b).

The results are compared with the ΛCDM model [10,11]. The first model aligns closely with ΛCDM, as shown in Fig. 3(a), while Fig. 3(b) shows approximate agreement. Overall, the findings are consistent with ΛCDM across all datasets considered (CC, Pantheon$^+$SH0ES, CC+Pantheon$^+$SH0ES, and CC+Pantheon$^+$SH0ES+BAO) [34–41,47].

Fig. 4 shows the distance modulus $\mu(z)$ for different $f(R,T)$ gravity models, compared with the Pantheon$^+$SH0ES sample of 1701 light curves from 1550 supernovae. The model demonstrates good agreement with all datasets considered: CC, Pantheon$^+$SH0ES, CC+Pantheon$^+$SH0ES, and CC+Pantheon$^+$SH0ES+BAO [34–41,47].

Fig. 4 show the distance modulus with model curves fitted to the Pantheon$^+$SH0ES datasets and error bars. The model consistently fits the Pantheon$^+$SH0ES observations across all dataset combinations (CC, Pantheon$^+$SH0ES, CC+Pantheon$^+$SH0ES, and CC+Pantheon$^+$SH0ES+ BAO) [34–41,47].

## 5. Cosmological parameters

In this section, we examine how important variables evolve over time in $f(R,T)$ gravity models, using observational data to guide our analysis.

### 5.1. Deceleration parameter

The deceleration parameter $q(z)$ is a fundamental cosmological quantity that characterizes the expansion dynamics of the Universe; it determines whether cosmic expansion is accelerating or decelerating. However, the universe's shift from a decelerating to an accelerating phase marks a pivotal development in modern cosmology, signifying the influence of a repulsive component, commonly referred to as dark energy, which presently dominates the expansion of the Universe [68–72].

The deceleration parameter is directly related to the Hubble parameter, $H(z)$, which quantifies the expansion rate at a given redshift. This relationship enables rigorous testing of cosmological models and enhances understanding of dark energy and the long-term evolution of the Universe. The deceleration parameter is expressed in terms of redshift $z$ as follows [73,74]

$$q(z) = -1 + (1+z)\frac{H'(z)}{H(z)}, \quad (5.1)$$

where the prime indicates differentiation with respect to redshift $z$.

Fig. 5(a) and (b), shows that the deceleration parameter approaches $q \approx 0.5$ at high redshift, consistent with the ΛCDM model and indicative of a matter-dominated era characterized by decelerating expansion [10,11]. As the Universe evolves, the deceleration parameter changes sign near the transition redshifts $z_{tr}^{CC} = 0.62$, $z_{tr}^{Pan} = 0.81$, $z_{tr}^{CC+Pan} = 0.84$, and $z_{tr}^{CC+Pan+BAO} = 0.75$ for the first model (Fig. 5(a)), and $z_{tr}^{CC} = 0.35$, $z_{tr}^{Pan} = 0.83$, $z_{tr}^{CC+Pan} = 0.81$, and $z_{tr}^{CC+Pan+BAO} = 0.75$ for the second model (Fig. 5(b)), indicating the onset of accelerated expansion. Over time, the deceleration parameter $q$ is expected to approach $q = -1$, which signifies a transition to a de Sitter phase with constant, exponential expansion, where dark energy becomes increasingly dominant relative to matter and radiation. At the present epoch ($z = 0$), the deceleration parameter values are $q_0^{CC} = -0.33$, $q_0^{Pan} = -0.63$, $q_0^{CC+Pan} = -0.62$, and $q_0^{CC+Pan+BAO} = -0.59$ for the first case (Fig. 5(a)), and $q_0^{CC} = -0.52$, $q_0^{Pan} = -0.63$, $q_0^{CC+Pan} = -0.65$, and $q_0^{CC+Pan+BAO} = -0.62$ for the second case (Fig. 5(b)). These results closely match current observations, supporting the models and reinforcing the scenario of a Universe dominated by a positive cosmological constant or dark energy [73,74].

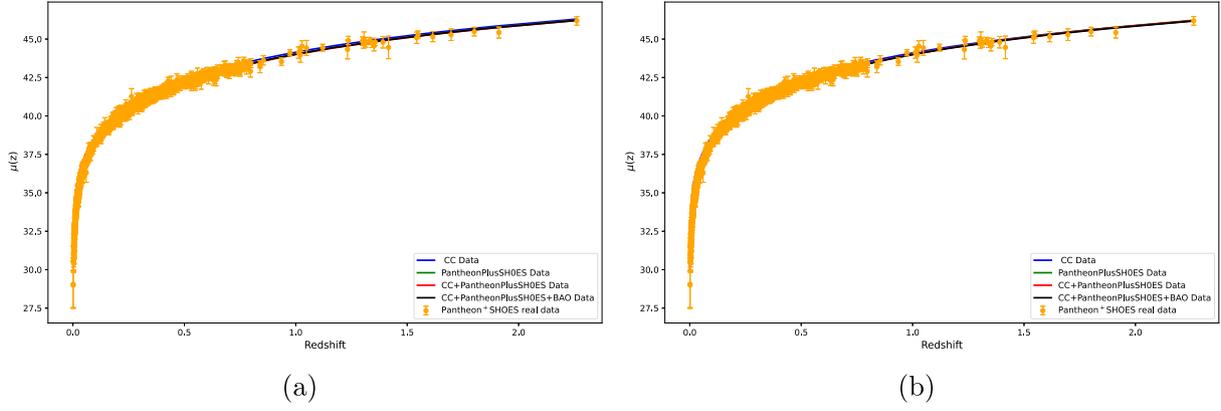

**Fig. 4.** Evolution of the distance modulus $\mu(z)$ for different value of $n$: (a) $n = 1$ and (b) $n \neq 1$, constrained with various observational datasets (CC, Pantheon$^+$SH0ES, CC + Pantheon$^+$SH0ES, and CC + Pantheon$^+$SH0ES + BAO).

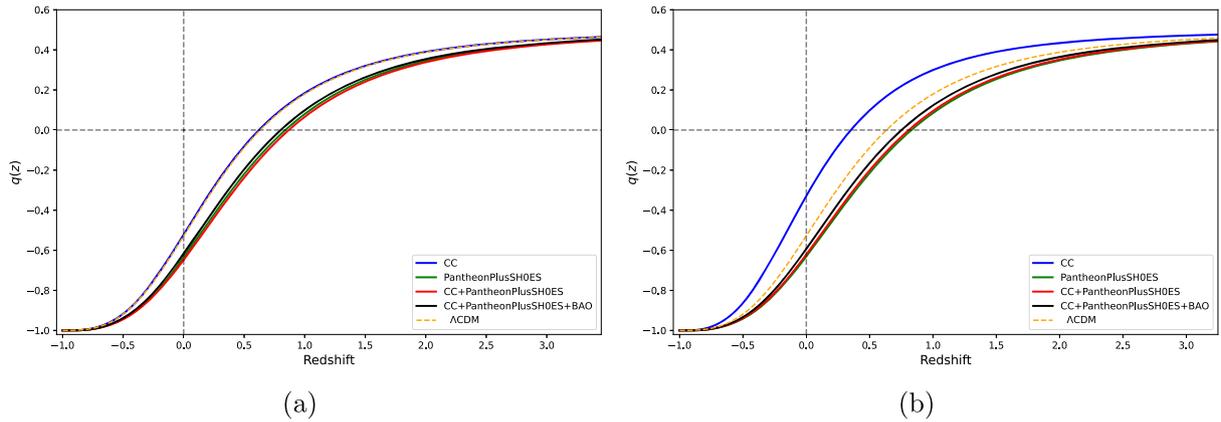

**Fig. 5.** Evolution of the jerk parameter j(z) for different values of $n$: (a) $n = 1$ and (b) $n \neq 1$, constrained using various observational datasets (CC, Pantheon$^+$SH0ES, CC + Pantheon$^+$SH0ES, and CC + Pantheon$^+$SH0ES + BAO). The purple dashed curve represents the $\Lambda$CDM prediction with $\Omega_\Lambda = 0.7$ and $\Omega_m = 0.3$.

### 5.2. Jerk parameter

The jerk parameter $j(z)$ as a function of redshift $z$ is given by [68–74]:

$$j(z) = 1 - 2\frac{1+z}{H(z)}H'(z) + \frac{(1+z)^2}{H(z)^2}(H'(z))^2 + \frac{(1+z)^2}{H(z)}H''(z) \quad (5.2)$$

Fig. 6(a) and (b) illustrate the evolution of the jerk parameter as a function of redshift for the first and second case, respectively. The current jerk parameter values are approximately $j_0 \approx 1.00001^{+0.000002}_{-0.000001}$ and $j_0 \approx 1.0001^{+0.00002}_{-0.00001}$ for the both cases and datasets, aligning with recent cosmographic studies. At both low and high redshift, the jerk parameter converges to one, as predicted by the standard $\Lambda$CDM model [10,11].

### 5.3. Snap parameter

The snap parameter $s(z)$ characterizes the fourth derivative of the scale factor and provides a higher-order diagnostic of cosmic expansion, extending analysis beyond the deceleration and jerk parameters. The general expression for $s(z)$ is formulated in terms of the jerk $j(z)$ and deceleration $q(z)$ parameters, both dependent on $H(z)$ and its derivatives [68–72]. The snap parameter is given by

$$s(z) = j(z)'(1+z) + j(z)[1 - 2q(z)] \quad (5.3)$$

Here, the prime denotes differentiation with respect to redshift $z$. Fig. 7 illustrates the evolution of the snap parameter $s(z)$ across the redshift range. At low redshift ($z \approx 0$), all datasets yield $s(z)$ values close to those predicted by the $\Lambda$CDM model, indicating consistency with a dark energy-dominated, accelerating Universe [10,11]. However, at higher redshifts, the evolution of $s(z)$ exhibits a distinct redshift dependence compared to the standard $\Lambda$CDM expectation.

For future epochs ($z < 0$), $s(z)$ attains positive values, reflecting the dominance of dark energy-driven acceleration. At the present epoch ($z \approx 0$), the snap parameter is negative, indicating a deviation from the $\Lambda$CDM prediction and highlighting the influence of $f(R,T)$ modifications [10,11]. As redshift increases, $s(z)$ remains negative across the observable range but gradually approaches a constant value for $z \geq 2.5$. This asymptotic behavior suggests that at high redshift, the expansion history becomes largely insensitive to the specifics of the modified gravity model and converges toward a matter-dominated evolution [68–72]. Overall, the trend indicates a positive snap in future epochs, a negative snap at the present and intermediate redshifts, and a constant asymptotic value at high redshift, reflecting the transition from dark energy-dominated accelerated expansion to standard matter-driven dynamics in the early Universe.

### 5.4. Equation of state parameter

The Equation of State (EoS) parameter, $\omega$, is a key metric for distinguishing different epochs of cosmic expansion, its value provides insight into the fundamental properties of the Universe components. Specifically, $\omega = 1$ corresponds to a stiff fluid with rapid expansion, $\omega = 1/3$ characterizes the radiation-dominated era, $\omega = 0$ indicates the matter-dominated phase, and the current accelerating expansion is defined by $\omega < -1/3$, which signifies the dominance of dark energy with negative pressure.

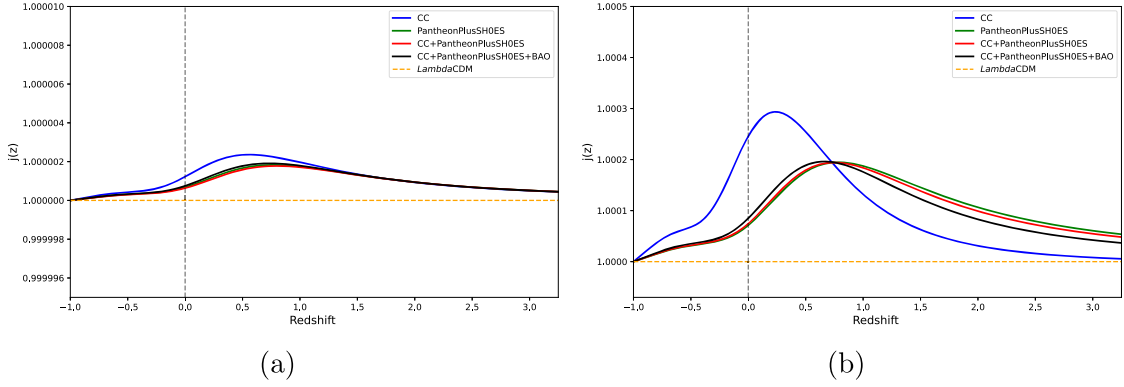

**Fig. 6.** Evolution of the jerk parameter $j(z)$ for different value of $n$: (a) $n = 1$ and (b) $n \neq 1$, each constrained by various observational datasets (CC, Pantheon$^+$SH0ES, CC + Pantheon$^+$SH0ES, and CC + Pantheon$^+$SH0ES + BAO). The purple dashed curve shows the $\Lambda$CDM prediction with $\Omega_\Lambda = 0.7$ and $\Omega_m = 0.3$.

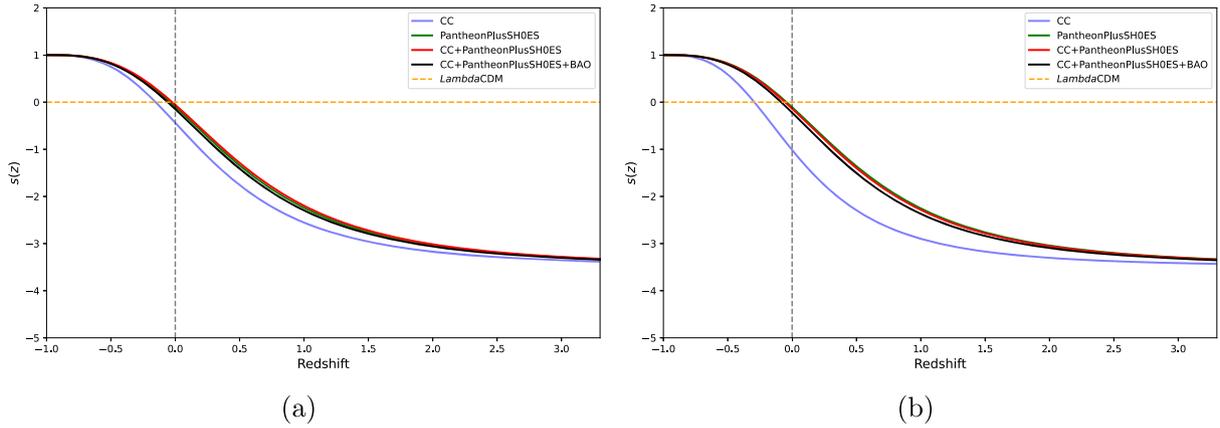

**Fig. 7.** The evolution of the snap parameter $s(z)$ for different values of $n$: (a) $n = 1$ and (b) $n \neq 1$, constrained using various observational datasets (CC, Pantheon$^+$SH0ES, CC + Pantheon$^+$SH0ES, and CC + Pantheon$^+$SH0ES + BAO). The purple dashed curve denotes the $\Lambda$CDM prediction with $\Omega_\Lambda = 0.7$ and $\Omega_m = 0.3$.

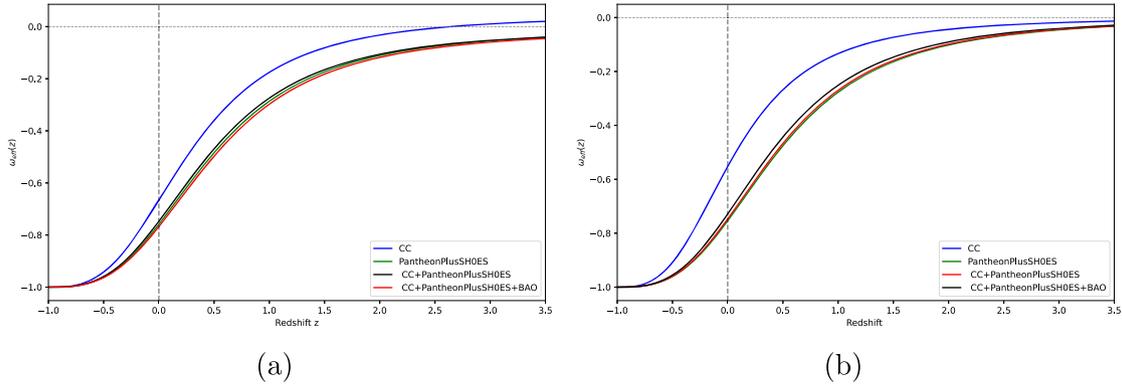

**Fig. 8.** The evolution of the total EoS parameter $\omega_{eff}$ for different value of $n$: (a) $n = 1$ and (b) $n \neq 1$, constrained with various observational datasets (CC, Pantheon$^+$SH0ES, CC + Pantheon$^+$SH0ES, and CC + Pantheon$^+$SH0ES + BAO).

Fig. 8 display the effective EoS parameter for both cases across various datasets (CC, Pantheon$^+$SH0ES, CC + Pantheon$^+$SH0ES, CC + Pantheon$^+$SH0ES + BAO) [34–41,47], showing that it remains below $-1/3$ at $z = 0$. For the first case Fig. 8(a), at $z = 0$, the range $-0.81 < \omega_{\text{eff}} < -0.64$ is closer to the $\Lambda$CDM value $\omega = -1$. In contrast, for the second case Fig. 8(b) at $z = 0$, the range $-0.75 < \omega_{\text{eff}} < -0.57$ is obtained for the different datasets, indicating a weaker acceleration [10,11]. Both models also exhibit a crossing of $\omega = -1$, suggesting that the phantom-like acceleration may result from quintessence-type dark energy or effective modified-gravity effects.

### 5.5. Energy conditions

This section evaluates the standard energy conditions to assess the physical viability of the cosmological models. The Null Energy Condition (NEC), Dominant Energy Condition (DEC), and Strong Energy Condition (SEC) are defined in terms of energy density $\rho$ and pressure $p$ as follows: NEC $\rho + p \geq 0$, DEC $\rho - p \geq 0$, and SEC $\rho + 3p \geq 0$. The functions $\rho(z)$ and $p(z)$ for two different cases, are used to evaluate the relevant combinations for each energy condition [87–89,89–96].

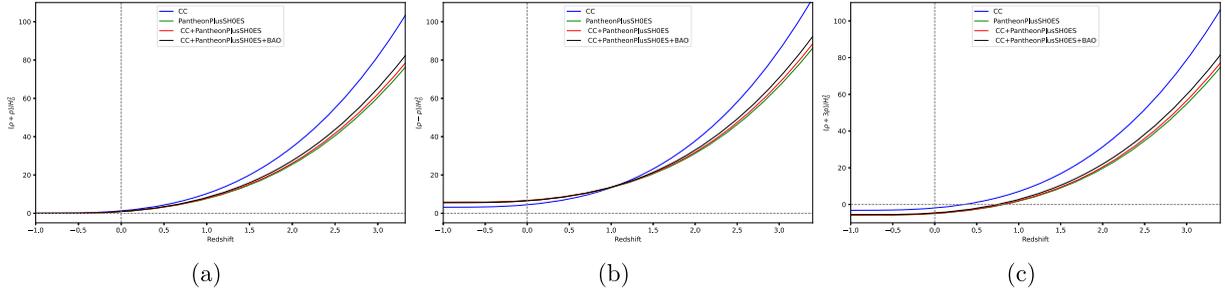

**Fig. 9.** The growth of energy condition with respect to redshift z for different observational data. (a) NEC $(\rho + p)/H_0^2$, (b) DEC $(\rho - p)/H_0^2$, and (c) SEC $(\rho + 3p)/H_0^2$ for $n \neq 1$.

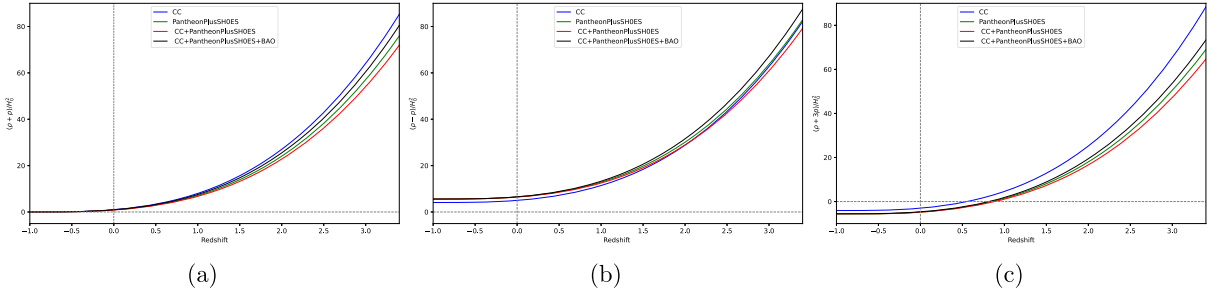

**Fig. 10.** The evolution of energy condition with respect to redshift z for different data. (a) NEC $(\rho + p)/H_0^2$, (b) DEC $(\rho - p)/H_0^2$, and (c) SEC $(\rho + 3p)/H_0^2$ for $n = 1$.

Figs. 10 and 9 present the evolution of energy conditions as functions of redshift for two cases, using the datasets CC, Pantheon$^+$SH0ES, CC + Pantheon$^+$SH0ES, and CC + Pantheon$^+$ SH0ES + BAO. Figs. 10(a) and 9(a) illustrate the evolution of $(\rho + p)/H_0^2$ with redshift for these datasets. In both cases, the null energy condition (NEC) is satisfied for all parameter values, as the curves remain non-negative across the redshift range. At low redshifts ($z \sim 0$), $(\rho + p)/H_0^2$ approaches zero, confirming that the NEC holds at present. At higher redshifts ($z > 1$), the values increase to large positive numbers. Figs. 10(b) and 9(b) depict $(\rho - p)/H_0^2$ as a function of redshift, confirming that the dominant energy condition (DEC), $(\rho - p) \geq 0$, is satisfied throughout the redshift range. At low redshifts, $(\rho - p)/H_0^2$ begins with small positive values and increases with increasing $z$. Figs. 10(c) and 9(c) display $(\rho + 3p)/H_0^2$ as a function of redshift, allowing examination of the strong energy condition (SEC) violation, $(\rho + 3p) < 0$. The results indicate that $(\rho + 3p)/H_0^2$ remains negative for negative values of $z$, implying SEC violation in the present and future. At higher redshifts ($z > 0$), $(\rho + 3p)/H_0^2$ increases to positive values, reflecting the dominance of matter and radiation in the early Universe. Overall, Figs. 10 and 9 demonstrate that $\rho + p \geq 0$ and $\rho + 3p \geq 0$, indicating that both the NEC and SEC are satisfied at present ($z = 0$) and in the future ($z \to -1$). The figures also show that $\rho - p \geq 0$, suggesting that the DEC is violated for both models and all datasets (CC, Pantheon$^+$SH0ES, CC + Pantheon$^+$SH0ES, and CC + Pantheon$^+$SH0ES + BAO) [34–41,47].

## 6. Conclusion

This study investigates the cosmological implications of modified gravity forms of $f(R, T) = R + \alpha T^n$, where $\alpha$ and $n$ are free parameters, by considering the two different cases: the general and nonlinear $n \neq 1$, and the linear one where $n = 1$. The modified Friedmann equations are used to derive the background evolution of the Hubble parameter $H(z)$, and the models are compared with multiple observational datasets, including cosmic chronometers (CC), the updated Pantheon$^+$SH0ES compilation, baryon acoustic oscillations (BAO), and their combinations. Parameter constraints are obtained through $\chi^2$ minimization and MCMC sampling [32], using the emcee Python library [33]. For the first case $n = 1$, the Hubble constant is found to be in the range $H_0 \simeq 62.5\text{-}76.7, \text{km}, \text{s}^{-1}, \text{Mpc}^{-1}$, overlapping with both the Planck 2018 and SH0ES measurements. The coupling $\alpha$ remains consistent with zero at $2\sigma$, indicating no significant deviation from general relativity. In the second case, $n \neq 1$, the constraints are fully compatible with $\Lambda$CDM, yielding $H_0 \simeq 67\text{-}68, \text{km}, \text{s}^{-1}, \text{Mpc}^{-1}$, $\Omega_m \simeq 0.31\text{-}0.55$, and $n \approx 0.5$ [10,11]. Both cases reproduce the late-time acceleration of the Universe and remain observationally indistinguishable from $\Lambda$CDM, while allowing small deviations parameterized by $\alpha$ and $n$; the best-fit values are summarized in Tables 3-2. Thus, dynamical analysis supports these results. In addition, the deceleration parameter $q(z)$ Fig. 5 exhibits a transition from positive to negative values, with transition redshifts $0.62 < z_{tr} < 0.81$ for the first model and $0.35 < z_{tr} < 0.83$ for the second,These intervals are consistent with observational estimates of the onset of cosmic acceleration, at the present epoch, $q(0) < 0$, confirming the accelerating expansion of the Universe. Moreover, The jerk parameter remains close to $j(z) \approx 1$ Fig. 6, in agreement with $\Lambda$CDM, particularly for the linear model where $j(z) \approx 1.0000001$ Fig. 6(a). Furthermore, the effective equation-of-state parameter $\omega_{\text{eff}}(z)$ Fig. 8 remains negative throughout cosmic evolution, consistent with an accelerating Universe. The snap parameter $s(z)$ Fig. 7 is positive in the future, negative at low to intermediate redshifts, and asymptotically constant at high redshift, reflecting the transition from dark energy to matter domination. The analysis of the energy conditions shows that both the NEC and DEC are satisfied throughout cosmic history, confirming the absence of exotic or phantom-like matter. The strong energy condition (SEC) is violated at low redshifts ($z \leq 0.7$) but satisfied at higher redshifts ($z \geq 0.7$). This behavior explains the transition from the early decelerating matter-dominated phase to the present accelerated expansion, consistent with observational evidence of the cosmic acceleration epoch. In summary, the results demonstrate that the considered $f(R, T)$ gravity models are consistent with current observational data, reproduce the late-time acceleration, and provide a viable alternative framework to $\Lambda$CDM while allowing for small but testable deviations [10,11].

**Data availability**

The data that has been used is confidential.

**Declaration of competing interest**

The authors declare that they have no known competing financial interests or personal relationships that could have appeared to influence the work reported in this paper.

**Acknowledgements**

the authors would like to thank Bouzid Manaut for his discussions and scientific help. The authors would like to thank the anonymous reviewers for their helpful comments, remarks, and suggestions, which significantly improved the overall manuscript.